\documentclass[12pt]{article}

\newcommand{\be}{\begin{equation}}
\newcommand{\ee}{\end{equation}}
\newcommand{\bea}{\begin{eqnarray}}
\newcommand{\eea}{\end{eqnarray}}

\newcommand{\bit}{\begin{itemize}}
\newcommand{\eit}{\end{itemize}}
\newcommand{\no}{\noindent}

\begin{document}
{\sf \title{Calabi-Yau 3-folds from 2-folds}
\author{Ansar Fayyazuddin\footnote{email: Ansar\_ Fayyazuddin@baruch.cuny.edu} }
\maketitle
\begin{center}
\vspace{-1cm}
{\it $^1$ Department of Natural Sciences, Baruch College, \\
City University of New York, New York, NY}
\end{center}

\begin{abstract}
We consider type IIA string theory on a Calabi-Yau 2-fold with D6-branes wrapping 2-cycles in the 2-fold.  We find a complete set of conditions on the supergravity solution for any given wrapped brane configuration in terms of SU(2) structures.  We reduce the problem of finding a supergravity solution for the wrapped branes to finding a harmonic function on R$\times$CY$_2$.   We then lift this solution to 11-dimensions as a product of R$^{(4.1)}$ and a Calabi-Yau 3-fold.  We show how the metric on the 3-fold is determined in terms of the wrapped brane solution.  We write down the distinguished (3,0) form and the K{\"a}hler form of the 3-fold in terms of structures defined on the base 2-d complex manifold.  We discuss the topology of the 3-fold in terms of the D6-branes and the underlying 2-fold.  We show that in addition to the non-trivial cycles inherited from the underlying 2-fold there are $N-1$ new 2-cycles.  We construct closed (1,1) forms corresponding to these new cycles.  We also display some explicit examples.  One of our examples is that of D6-branes wrapping the 2-cycle in an A$_1$ ALE space, the resulting 3-fold has $h^{(1,1)}=N$, where $N$ is the number of D6-branes.      
\end{abstract}

\vspace{-16cm}
\begin{flushright}
BCCUNY-HEP /07-01 \\
hep-th/0702135
\end{flushright}

\thispagestyle{empty}

\newpage

\tableofcontents

\section{Introduction}
In two recent papers \cite{fh, fh2} a framework was developed to find constraints on supergravity solutions for wrapped branes and to give these constraints, traditionally found through complicated Killing spinor techniques, a transparent physical meaning.  The methods have the additional advantage of being computationally efficient.  This paper is an application of these techniques to the problem of D6-branes wrapping supersymmetric 2-d submanifolds in Calabi-Yau 2-folds in 10-d type IIA supergravity.  

Aside from the intrinsic interest of understanding wrapped brane supergravity solutions, there is an additional bonus when the wrapped branes are D6-branes - their lift to 11-dimensions is purely geometric.  Thus the wrapped brane solutions can be used to study supersymmetric geometries providing, one hopes, some useful insights.  We will study the problem from both the wrapped brane perspective as well as its lift to the geometry of a product of 5-d Minkowski space and a Calabi-Yau 3-fold.  We will focus on this Calabi-Yau 3-fold and relate the geometric and topological structures that arise on the 3-fold to those of the underlying 2-fold.   An intricate relationship between the 2- and 3-folds emerges.   

The problem we study is one of many possible scenarios in which D6-branes yield interesting geometries through their M-theory lift.  This rich structure of wrapped D6-branes in various contexts is discussed in depth in \cite{gm}.  Providing a detailed exploration of these different scenarios is another motivation for this paper.  In addition to \cite{gm} the problem of wrapped D6-branes and their lift to Calabi-Yau and G2-holonomy manifolds was explored from a different perspective in \cite{enunez}, where several metrics were explicitly constructed.

This paper is structured as follows.  In section 2 we explain in more detail the problem studied here, including some background on Calabi-Yau 2-folds.  In section 3 we describe our methods based on \cite{fh, fh2} for deducing constraints on supergravity solutions for wrapped branes.  In sections 4 and 5 we apply these methods to the problem and make some observations concerning the solution.  In section 6 we show how the Calabi-Yau 3-fold arises from the lift to 11-d and give a detailed description of its geometry and topology.  In section 7 some explicit examples are discussed, and finally in section 8 we conclude with a summary of our results.

\section{Calabi-Yau 2-folds and wrapped D6-branes}
The methods employed in this paper were explained in some detail in \cite{fh, fh2}\footnote{See also \cite{martucci} where D-brane calibrations are developed independently in the context of space-filling D-branes.}.  We briefly review them here before proceeding to their applications to the problem at hand.

Our goal is to give a complete description of the features of supergravity solutions of wrapped D6-branes in low-energy type IIA string theory compactified on a Calabi-Yau 2-fold.  It is immaterial for what follows whether the Calabi-Yau 2-fold is compact or not.  We will sometimes refer to the 2-fold in shorthand as a K3. 

We will work simultaneously at three different levels of description.  The first level is that of type IIA string theory at low energies (i.e. describable by supergravity) on R$^{(5,1)}\times$K3.  This background is invariant under 16 real supercharges.  At the next level we have D6-branes wrapped on 2-cycles of the K3.  This is the actual configuration that we seek to describe in supergravity. The solution should preserve 8 real supersymmetries.  At the third level we introduce probe branes into the D6 supergravity background.  The probe branes are introduced to help us determine the wrapped D6 solution, they will not modify the supergravity background.  We will pick these probe branes so that they have an interpretation as objects in the flat part of the D6-brane worldvolume theory.

More concretely, consider type IIA string theory on a K3.  The supergravity solution consists of a constant dilaton and a metric given by:
\be
ds^2 = \eta_{\mu\nu}dx^\mu dx^\nu + dy^2 + 2h_{M{\bar N}}dz^M dz^{\bar N}. \label{k3}
\ee
Here $h_{M{\bar N}}$ is the metric on the K3 with local complex coordinates $z^M$, $M=1,2$.  The remaining space-time is six dimensional flat Minkowski space.  For later convenience we have split this space into 5-d Minkowski space with coordinates $x^\mu, \mu =0,1,...,4$ and a sixth coordinate $y$.  

The K3 has non-trivial 2-cycles on which we can wrap branes.  K3 is a complex manifold under incompatible complex structures.  These inequivalent complex structures can be rotated into each other by an SU(2) rotation.  This is a reflection of the fact that K3 manifolds are hyper-K{\" a}hler.  Any given supersymmetric 2-cycle in the K3 is holomorphic with respect to a particular complex structure but not with respect to others.  In what follows we will pick a complex structure on the K3 so that the 2-cycle we are wrapping D6-branes on is holomorphic.  

The K3 has two distinguished 2-forms: the (1,1) K{\" a}hler form and a (2,0) holomorphic form.  The (p,q) label of forms is defined with respect to a fixed complex structure- in our case one that is compatible with the wrapped 2-cycle being holomorphic.  Since the cycle on which the brane is wrapped is holomorphic, the (2,0) form vanishes on the cycle while the pullback of the K{\" a}hler form is its volume form.  This fact is often summarized by saying that holomorphic cycles are calibrated by the K{\" a}hler form.  These calibrated cycles are supersymmetric in the sense that branes wrapping them will preserve a fraction of the supersymmetry.  

The problem we seek to solve is: how can one characterize supergravity solutions for wrapped D6-branes.  The wrapped D6-branes have two of their directions inside of the K3 and are localized in one direction transverse to the K3.  We will denote by $y$ the coordinate in which the D6-brane is localized transverse to the K3.  The D6-brane is Poincare invariant in the worldvolume directions transverse to the K3.  Since the D6-branes under investigation wrap a holomorphic cycle, it is reasonable to assume that they preserve the complex structure of the underlying K3.  Therefore an ansatz compatible with these considerations is:
\be
ds^2_{10} =  H_1^2\eta_{\mu\nu}dx^\mu dx^\nu + H^2_2dy^2 + 2g_{M{\bar N}}dz^M dz^{\bar N}, \label{g10}
\ee
where $H_1, H_2$ are independent of $x^\mu$ due to Poincare invariance in these directions.   In addition to the metric we will assume that the dilaton and RR 1-form potential are allowed to vary.  We will set to zero the NSNS 2-form and R-R 3-form explicitly.  The reason we set these two fields to zero is because we wish to consider only those configurations that lift to pure geometry in 11-d.

This is our ansatz.  In the next section we determine the constraints on it by probing it with branes.

\section{Probing the background with branes}

Into this, yet to be completely determined supergravity background, we now introduce brane probes and study them as objects in the worldvolume field theory in the flat part of the D6-brane.  The D6-brane worldvolume field theory is five dimensional.  Among the different objects that can exist in this theory are objects that come from branes intersecting the D6-brane.  The intersections we introduce are all supersymmetric and are therefore static objects in the worldvolume theory.  

The worldvolume theory on a {\it static} probe p-brane can be written as:
\be
S = m\int dt
\ee
where $m$ is the effective mass of the brane:
\be
m = \int {\cal L}d\sigma^1...d\sigma^p.
\ee
$\cal L$ is the worldvolume Lagrange density of the brane and the $\sigma^i$ are spatial worldvolume coordinates.  If the brane has $q$ directions along $x^\mu$ then the intersection appears as a $q$-brane in the worldvolume of the D6-brane.  The tension of the $q$-brane is given by:
\be
T_q = \int_{\Sigma_{p-q}} {\cal L}d\sigma^{q+1}...d\sigma^p.
\ee
where $\Sigma_{p-q}$ is the $(p-q)$-dimensional submanifold that the $p$-brane wraps inside the K3.
We will show that there is a $(p-q)$-form which when integrated over the submanifold computes the tension:
\be
T_q = \int \phi |_{\Sigma_{p-q}}.
\ee
We will take this to be an indication that $\phi$ calibrates the tension and, like calibrations, require that it is closed:
\be
d_4\phi = 0
\ee 
where $d_4$ is the exterior derivative on K3.  The above conditions will provide us with the torsion class constraints for the geometry (\ref{g10}).  The derivation provides a physical reason for the torsion class constraint - it is the condition that the calibration is closed.

The Lagrange density appearing in the above expression depends on the type of brane being used as a probe.  For D-branes the Lagrangian can be written as:
\be
{\cal L} _{Dp}= T_p e^{-\phi}\sqrt{\det G}.
\ee
Here $T_p$ is the 10-d tension of the Dp-brane, $\phi$ is the dilaton and $G$ is the pullback of the space-time metric onto the worldvolume of the brane.  For NS5-branes the Lagrange density is given by:
\be
{\cal L}_{NS5} = T_{NS5}e^{-2\phi}\sqrt{\det G}\sqrt{1+e^{2\phi}a_aa_bG^{ab}} \label{ns5}
\ee
where $a$ is the pullback of the RR 1-form $A$ onto the worldvolume of the NS5-brane.

Now that the complex structure is fixed by requiring that the wrapped cycle is holomorphic, we can consider other submanifolds which may intersect the D6-brane.  A holomorphic submanifold will intersect our chosen holomorphic cycle at either a point or in 1-complex dimension.  A Special Lagrangian submanifold is one on which the K{\" a}hler form vanishes and the pullback of our distinguished (2,0) form on to it is its volume form (upto a phase).  Such a Special Lagrangian submanifold intersects our holomorphic cycle along 1-real dimension.  The dimensions of the intersection are important pieces of information for us.

The particular calibration deduced from the Lagrangian depends on the brane configuration considered.  We will now systematically examine different objects that appear in the D6-brane worldvolume while preserving some fraction of the supersymmetry.  Since supersymmetry is a local condition it is enough to think about flat intersecting branes to determine what the supersymmetric configurations are.  

\begin{itemize}

\item D0-branes.  In this case there are no supersymmetric intersections.  

\item D2-branes.  To preserve supersymmetry  D2-branes have to lie either entirely inside the D6-brane or completely transverse to it.  In other words the intersection has to be either 2-dimensional or 0-dimensional.  Since our background D6-branes wrap a holomorphic 2-cycle in the K3, the D2-branes must either wrap a holomorphic 2- or 0-submanifold.

\item D4-branes.  Supersymmetric D4-branes can either intersect D6-branes along $3$ or $1$ space directions.  In our case these two possibilities correspond to D4-branes wrapped on Special Lagrangian 2-submanifolds  and localized in $y$ or wrapped on holomorphic 2-submanifolds while also stretched in the $y$ direction between D6-branes separated along $y$. 

\item D6-branes.  For the intersection to lie within the flat part of the worldvolume the probe D6-branes have to wrap a holomorphic 2-fold - the entire K3.  

\item F1-strings.  They can stretch between D6-branes separated along $y$.  

\item NS5-branes.  NS5-branes can intersect the wrapped D6-branes along either 3 or 5 space directions.  This means that to preserve supersymmetry the probe NS5-brane either has (3+1) directions along $x^\mu$ and is wrapped on a holomorphic submanifold, or the NS5-brane has (2+1) directions along $x^\mu$, is stretched along $y$ and is wrapped on a special Lagrangian 2-submanifold.  
\end{itemize}

A simple argument shows that the volume forms on supersymmetric cycles on $\cal M$ are given by the generalizations of the calibration on the K3 to the appropriate object in the geometry of (\ref{g10}).  The argument is as follows.  The amount of supersymmetry preserved is accurately measured in the probe approximation.  In the probe approximation we find that the volume form on supersymmetric submanifolds in K3 
are either given by the K{\" a}hler form on K3 for holomorphic submanifolds or by the real part of the (2,0) form for special Lagrangian submanifolds (with an appropriate phase choice for the form).  The supersymmetry conditions rely only on how the gamma matrices with tangent space indices act on the spinor, which continue to hold true in the full geometry of (\ref{g10}).  Therefore when we go from tangent space to curved space indices we simply replace the K{\" a}hler form on K3 with the K{\" a}hler form on ${\cal M}$ and similarly the (2,0) form on K3 is replaced by the (2,0) form on ${\cal M}$.  By ${\cal M}$ we mean the part of the space time (\ref{g10}) that has coordinates $z^M$ - i.e. what K3 is replaced by in the full geometry.  

The above identification of volume forms on supersymmetric submanifolds as either the pullback of $J$ or ${\sf Re} (e^{i\alpha}\Omega)$ (for some appropriate constant phase $\alpha$), will be important for our analysis as we begin to analyze the various configurations listed above.  Let us consider one by one the above supersymmetric configurations. 

Consider first D2 branes transverse to ${\cal M}$ and completely inside the D6-brane.   The D2-branes appear as 2-branes in the flat part of the worldvolume theory of the D6-brane.  The action of the D2-brane is:
\be
S = T_2\int e^{-\phi}H_1^3 dt\wedge dx^1\wedge dx^2
\ee
The tension of the 2-brane is then:
\be
T = T_2e^{-\phi}H_1^3
\ee
According to our general scheme then $e^{-\phi}H_1^3$ is a calibration for the tension of the 2-brane and, therefore:
\be
d_4(e^{-\phi}H_1^3)=0
\ee
or 
\be
e^{-\phi} = H_1^{-3}\label{dilaton}
\ee
where the dilaton $\phi$ is defined so that its asymptotic value as $y\rightarrow \infty$ is $0$.  In principle there can be a function of $y$ multiplying the right hand side of the above expression.  However, such a factor will result in the M-theory lift of the geometry to be a warped product of Minkowski space and a curved manifold rather than a simple product.  So we can conclude that the above identification is correct as we expect to lift to Minkowski space times a Calabi-Yau 3-fold.

Next consider a D2-brane wrapping a holomorphic 2-submanifold $\Sigma_2$ in ${\cal M}$.  In this case the intersection is 0 dimensional in the flat part of the D6-brane worldvolume.  The action for the D2-brane is given by:
\be
S= T_2\int_{R\times \Sigma_2} e^{-\phi} H_1 dt\wedge J
\ee
where the factor of $H_1$ is from the $G_{00}$ component of the metric, $J$ is the K{\" a}hler form which when pulled back to $\Sigma_2$ is the volume form on it.  The mass of the 0-brane in the D6-brane worldvolume theory is given by:
\be
m =T_2 \int_{\Sigma_2} e^{-\phi}H_1 J = T_2\int_{\Sigma_2} H_1^{-2} J \label{d2}
\ee
here we made use of (\ref{dilaton}) in the second equality.  From this we deduce that $H_1^{-2} J$ is a calibration for the mass of the 0-brane and therefore:
\be
d_4(H_1^{-2} J) =0 \label{kahler}
\ee
Thus the rescaled metric $H_1^{-2} g$ is K{\" a}hler.

Consider a D4-brane wrapping a special Lagrangian submanifold $\Sigma_2$ in $\cal M$ with the remaining directions of the D4-brane along $x^\mu$.  This appears as a 2-brane inside the D6-brane worldvolume theory.  The action of the D4-brane is:
\be
S = T_4\int_{R^{1,2}\times \Sigma_2}e^{-\phi}H_1^{3} dt\wedge dx^1\wedge dx^2\wedge \Omega
\ee
We have written the volume form on $\Sigma_2$ as $\Omega$ as a short hand for ${\sf Re}e^{i\alpha}\Omega$.  Since the configuration is supersymmetric independent of the constant phase $\alpha$ any conditions we derive are equally valid for $\Omega$.  From the above expression we can read off the tension of the 2-brane in the D6-brane worldvolume theory:
\be
T = T_4\int_{\Sigma_2}e^{-\phi}H_1^{3} \Omega.
\ee
This implies that the calibrating form for the tension is $e^{-\phi}H_1^{3} \Omega= \Omega$, from this it follows that:
\be
d_4\Omega =0.
\ee
Thus $\Omega$ is a holomorphic (2,0) form. 

D4-branes can also be stretched along the $y$ direction as long as they wrap a holomorphic 0- or 1-submanifold in the K3.  The wrapped D6-branes can be separated along $y$, the D4-branes then end on the D6-branes.  Taking the 0-cycle case first, the D4-brane appears as a 3-brane in the flat part of the D6-brane world volume theory.  
The action of the D4-brane is:
\be
S = T_4\int e^{-\phi}H_1^{4}H_2 dt\wedge dx^1\wedge dx^2\wedge dx^3\wedge dy \label{d4}
\ee
The tension of the 3-brane is given by:
\be
T = T_4\int e^{-\phi}H_1^{4}H_2 dy
\ee
which implies the calibration condition:
\be
0 = d_4(e^{-\phi}H_1^{4}H_2 dy) = d_4(H_1H_2 dy). 
\ee
From this condition it follows that:
\be
H_1 = H_2^{-1}. \label{h}
\ee
In principle one could include an additional arbitrary function of $y$ multiplying the right hand side, say, of the above equation.  However, such a function can be absorbed into a redefined $y$ coordinate.  
We turn now to the case of a D4-brane stretched along $y$ and also wrapping a holomorphic 1-submanifolds in the K3.  This appears as a string in the worldvolume theory of the D6-brane.  The action of the D4-brane is:
\be
S = T_4\int_{R^{(1,1)}\times \Sigma_2\times R} e^{-\phi}H_1^{2}H_2 dt\wedge dx^1\wedge J\wedge dy.
\ee
The tension of the string tin the D6-brane worldvolume theory is then given by:
\be
T = T_4\int_{\Sigma_2\times R} e^{-\phi}H_1^{2}H_2 dt\wedge dx^1\wedge J\wedge dy
\ee
which implies the calibration condition:
\be
0=d_4( e^{-\phi}H_1^{2}H_2 J\wedge dy) = d_4(H_1^{-2}J\wedge dy).
\ee
This equation is the same as our previous relation (\ref{kahler}), so it doesn't provide any new information.  However, it gives us confidence in our methods that the same relationship can be arrived at by a variety of means. 

Since supersymmetric D6-brane probes must wrap the entire K3, the calibrating 4-forms will be maximum rank forms on the K3, and therefore are trivially closed with respect to $d_4$.  So in this case we cannot extract any new information as far as the existence of closed forms is concerned.

We turn next to the NS-NS sector consisting of fundamental strings and NS5-branes.  If we introduce F1-string probes they can stretch between D6-branes  separated in the $y$ direction.  They appear as 0-branes in the worldvolume theory of the D6-branes.  The action of the F1 string is given by:
\be
S = T_{F1}\int H_1H_2 dt\wedge dy.
\ee
The mass of the 0-brane in the D6-brane worldvolume theory is given by:
\be
m = T_{F1}\int H_1H_2 dy
\ee
from which it follows that
\be
d_4(H_1H_2 dy) =0.
\ee
This is a relation that we discovered earlier in (\ref{h}) above.  

For the NS5-brane probes we use the expression for the Lagrangian given above (\ref{ns5}).  Consider the case of NS5-brane wrapped on a holomorphic submanifold of the K3.  The NS5-brane appears in the flat part of the D6-branes as a 3-brane.  In this case the action is:
\be
S = T_{NS5}\int_{R^{1,3}\times \Sigma_2} e^{-2\phi}H_1^{4}\sqrt{1+e^{2\phi}a_aa_bG^{ab}} dt\wedge dx^1\wedge dx^2\wedge dx^3\wedge J
\ee   
where $a_b$ is the pullback of $A$ onto the holomorphic 2-cycle.  Since the tension of the 3-brane is given by:
\be
T = T_{NS5}\int_{\Sigma_2} e^{-2\phi}H_1^{4}\sqrt{1+e^{2\phi}a_aa_bG^{ab}} J
\ee 
it is clear that the tension can be minimal only if $a_b=0$.  That is, the pullback of $A$ onto the cycle $\Sigma_2$ should vanish.  So we find that the tension is given by:
\be
T = T_{NS5}\int_{\Sigma_2} e^{-2\phi}H_1^{4} J = T_{NS5}\int_{\Sigma_2} H_1^{-2} J.
\ee 
This gives us no new calibration condition.

Finally, for NS5-branes wrapping special Lagrangian cycles stretched between D6-branes separated along $y$, the NS5-branes appear as 2-branes in the flat part of the D6-brane worldvolume theory.  The action of the NS5-brane is:
\be
S = T_{NS5}\int_{R^{1,2}\times \Sigma_2\times R} e^{-2\phi}H_1^{3}H_2\sqrt{1+e^{2\phi}a_aa_bG^{ab}} dt\wedge dx^1\wedge dx^2 \wedge \Omega\wedge dy
\ee   
from which the tension of the 2-brane in the D6-brane theory can be read off as:
\be
T = T_{NS5}\int_{\Sigma_2 \times R} e^{-2\phi}H_1^{3}H_2\sqrt{1+e^{2\phi}a_aa_bG^{ab}}\Omega\wedge dy.
\ee
As before the tension will be minimal only if $a_b$ vanish, that is the pullback of $A$ vanishes on the submanifold $\Sigma_2$.  In this case the tension is given by: 
\bea
T &=& T_{NS5}\int_{\Sigma_2 \times R} e^{-2\phi}H_1^{3}H_2\Omega\wedge dy \\
&=&T_{NS5}\int_{\Sigma_2 \times R} H_1^{-4}\Omega\wedge dy. \nonumber
\eea
It would appear that $d_4(H_1^{-4}\Omega)=0$.  This would not be good, since combined with our earlier condition $d_4\Omega=0$ would give us a very stringent constraint on $H_1$ allowing for relatively trivial solutions.  However it is important to realize that $H_1^{-4}\Omega$ is a calibration only on submanifolds on which $A$ has vanishing pullback.  In other words the condition is really that:
\be
d_4(H_1^{-4}(\Omega - \Omega_{MP}\frac{A^PA_N}{A_LA^L}dz^M \wedge dz^N)) =0.
\ee
This rather clumsy equation is expressed in a much more elegant form once we lift to 11-dimensions as we shall see.

\subsection{Summary and new notation}
In this subsection we collect the results just derived and introduce notation in which our results are somewhat more concisely expressed.  

In light of the above analysis, we re-write our ansatz for the metric (\ref{g10}) as follows:
\be
ds^2_{10} =  H^2\eta_{\mu\nu}dx^\mu dx^\nu + H^{-2}dy^2 + 2H^{2}g'_{M{\bar N}}dz^M dz^{\bar N}, \label{g10new}
\ee
Where we have defined $H = H_1 = H_2^{-1}$.  The metric $g'$ defined on ${\cal M}$ is related to $g$ through $g' = H^{-2} g$.  Our result (\ref{kahler}) shows that $g'$ is K{\" a}hler on $\cal M$.  We associate a K{\" a}hler form with this rescaled metric, $J' = H^{-2}J$ which is closed, but, consequently, $\Omega'$ defined with its usual relationship to $J'$ is not:
\bea
d_4 J' &=& 0\nonumber \\
d_4 (H^2\Omega') &=& 0 \\
J'\wedge J'& =&-\frac{1}{2}\Omega'\wedge \bar{\Omega}' \nonumber \label{jomega}
\eea   
The dilaton was determined in (\ref{dilaton}):
\be
e^{-\phi} = H^{-3} \label{dil}
\ee
expressed now in our new notation.

We have still not determined the R-R gauge field $A$.  We do this in the next section.

\section{Generalized calibrations and determining the R-R 1-form $A$}
We have used brane probes to determine a number of conditions on the supergravity solution for wrapped D6-branes.  In this section we will show how to determine $A$ by using generalized calibrations.  The essential idea behind generalized calibrations \cite{gencal} is that for BPS states, the charge and mass are equal to each other.  Generalized calibrations equate the Lagrange density of the brane, which is the mass density for static branes, to the tensor field electrically sourced by the brane.  In our case the D6-branes wrap a holomorphic 2-cycle of the K3.  A probe D6-brane has a worldvolume action given by:
\be
S = T_6 \int_{R^{1,4}\times \Sigma_2}e^{-\phi}H^{5}dt\wedge dx^1\wedge dx^2\wedge dx^3\wedge dx^4\wedge (H^2J')
\ee
We use this action to identify the 7-form that couples electrically to D6-branes: 
\bea
A_7 &=&e^{-\phi}H^{5}dt\wedge dx^1\wedge dx^2\wedge dx^3\wedge dx^4\wedge (H^2J') \nonumber \\
 &=& H^4dt\wedge dx^1\wedge dx^2\wedge dx^3\wedge dx^4\wedge J'
\eea
where we have simplified the expression using (\ref{dil}).  

The R-R 1-form is related to the above expression through:
\bea
F_2 &=& dA = *dA_7 \nonumber \\
&=&  *(- dt\wedge dx^1\wedge dx^2\wedge dx^3\wedge dx^4\wedge (d_4H^4\wedge J' +\partial_y (H^4J')\wedge dy)) \nonumber \\
&=& -i\partial_LH^{-4} dz^L\wedge dy + i\partial_{\bar M}H^{-4} dz^{\bar M}\wedge dy + \partial_y J'  \label{F}
\eea
Once we know $F_2$ we can in principle determine $A$.  We shall see that it is possible to write down a simple expression for it.  But first we impose the equations of motion for $A$.  We have by construction satisfied $d*F_2=0$, but still not required that $F_2$ is closed:
\be
0 = dF_2 = 2i\partial_L\partial_{\bar M}H^{-4} dz^L\wedge dz^{\bar M}\wedge dy + \partial_y^2J'\wedge dy
\ee
In components this equation states:
\be
0 = 2\partial_L\partial_{\bar M}H^{-4} + \partial_y^2g'_{L\bar M}. \label{eom}
\ee
Since $g'$ is K{\" a}hler, it can be expressed in terms of a K{\" a}hler potential $K$ as: $g'_{L\bar M} = \partial_L\partial_{\bar M}K(z,{\bar z},y)$.  If we express $g'$ in terms of $K$, (\ref{eom}) reduces to:
\be
0=\partial_L\partial_{\bar M}(2H^{-4} + \partial_y^2 K) 
\ee
from which one concludes that:
\be
2H^{-4} + \partial_y^2 K = f + \bar{f}
\ee
where $f$ is a holomorphic function of $z^M$.  Since $K$ is defined through $g'$ it is possible to shift $K$ by holomorphic and anti-holomorphic functions of $z^M$ without affecting their defining relationship.   This allows us to shift $f$.  We fix $f$ now by specifying the behavior as $y\rightarrow \infty$.  In this limit we are in a region far from the D6-branes, the metric should approach R$^{(1,4)}\times$K3, or $H\rightarrow 1, g'\rightarrow h$, where $h$ is the metric on the K3 in (\ref{k3}).  The behavior of $K$ is not determined by these conditions.  We choose $\partial_y^2K \rightarrow 0$ in this limit.  Thus without any loss of generality we pick:
 \be
 2H^{-4} + \partial_y^2 K = 2 \label{neweom}
 \ee
Using this equation it is now straightforward to write down a simple expression for $A$ so that it gives the correct field strength $F_2 = dA$ as given in (\ref{F}) :
\be
A = -\frac{i}{2}\partial_M\partial_y Kdz^M + \frac{i}{2}\partial_{\bar M}\partial_y Kdz^{\bar M}. \label{A}
\ee

We have found an expression for $A$ in terms of the K{\" a}hler potential $K$ for $g'$.  $K$ and $H$ satisfy equation (\ref{neweom}) which has to be imposed separately.

\section{Some further observations}
We have made some progress in writing down constraints on the supergravity solution of wrapped D6-branes.  At this point we would like to explore the consequences of some of the relations that we have written down.

We begin our analysis by trying to reconcile the three equations involving $J'$ and $ \Omega'$ collected in (\ref{jomega}).  We have already noted that $J'$ is closed and can be expressed in terms of a K{\"a}hler potential.  We have not yet properly analyzed the remaining two equations involving $\Omega'$.  The first of these states that $H^2\Omega'$ is $d_4$ closed.  Since $\Omega'$ is a (2,0) form on $\cal M$, we can conclude that $H^2\Omega'$ is holomorphic in $z^M$:
\be
\Omega' = 2H^{-2}f(z^M,y)dz^1\wedge dz^2
\ee
where $f$ is holomorphic in $z^M$, the factor of $2$ is introduced for later convenience.  We have allowed $f$ to have a dependence on $y$ for now.  We will see later that there is no such dependence.  Now using the third of the equations in (\ref{jomega}), we conclude that:
\be
\sqrt{\det{g'}} = H^{-4}|f|^2.
\ee
Consider the limit $y\rightarrow \infty$.  In this limit, far from the D6-branes, one should recover the product of K3 and Minkowski space geometry: $H\rightarrow 1$ and $g'\rightarrow h$ the metric on K3.  We will argue later that $\partial_y\Omega =0$, but if we assume this result now, we can conclude that
\be
|f|^2 = \sqrt{\det h}
\ee
where $h$ is the metric on the K3 as given in (\ref{k3}).  Finally we can conclude that
\be
H^{-4} = \frac{\sqrt{\det{g'}} }{\sqrt{\det h}} \label{h}
\ee 

That $H^{-4}$ is given in terms of the determinant of $g'$ means that our differential equation (\ref{neweom}) is non-linear.   This non-linearity is surprising.  The reason it's surprising is that we can have an arbitrary number of D6-branes distributed along $y$ without violating supersymmetry.  It is difficult to see how one can reconcile this fact with a non-linear equation.  A linear equation is easier to reconcile with this freedom to add D6-branes, since for linear equations one can superpose solutons.  We will explore one resolution to this problem.  The solution we propose is based on the fact that the complex submanifold that the D6-brane wraps is co-dimension one in the 2-fold.  We consider the possibility that (at least locally):
\be
K = K_0(z^M, z^{\bar N}) + K_1(y, F(z^M), {\bar F}(z^{\bar N})). \label{ansatz}
\ee
We have split the K{\" a}hler potential so that $K_0$ is independent of $y$, it is the K{\"a}hler potential for the K3 metric.  While $K_1$ depends on $y$, but the dependence on $z^M$ is only through a holomorphic function $F$ of these variables and its complex conjugate.  The reason for this peculiar restriction is that we have in mind that $F(z)=c$ defines the submanifold on which the D6-branes are wrapped.  The metric generated by $K$ can be written as:
\bea
g'_{M\bar N} = h_{M\bar N} + k_{M\bar N} \nonumber \\
h_{M\bar N} = \partial_M\partial_{\bar N}K_0 \\
k_{M\bar N} = \partial_M\partial_{\bar N}K_1 \nonumber.
\eea
Since $K_1$ depends on the coordinates on $\cal M$ through a single holomorphic function and its complex conjugate:
\bea
\det k &=& 0 \nonumber \\
\sqrt{\det g'} &=& \sqrt{\det h}(1 + h^{M\bar N}k_{M\bar N}). 
\eea
With this choice of K{\" a}hler potential, and our expression for $H$ given in (\ref{h}), we find:
\be
H^{-4} = (1 + h^{M\bar N}k_{M\bar N}).
\ee
Our non-linear differential equation (\ref{neweom}) now becomes:
\bea
0 &=& 2h^{M\bar N}k_{M\bar N} + \partial_y^2 K_1\nonumber \\
&=& 2h^{M\bar N}\partial_{M}\partial_{\bar N}K_1 + \partial_y^2 K_1 \label{diffk}
\eea
a linear one for $K_1$ due to our ansatz (\ref{ansatz}).  This linear equation states that $K_1$ is harmonic on K3.  We can transform this into an equation for $H$ by applying $\partial_y^2$ to the above equation:
\be
0 =  2h^{M\bar N}\partial_{M}\partial_{\bar N}H^{-4} + \partial_y^2 H^{-4}. \label{harmonic}
\ee
So, $H^{-4}$ is a harmonic function on K3.

There is an important consistency condition that must be satisfied if $K_1$ depends on the holomorphic coordinates only through a single holomorphic function (\ref{ansatz}).  The differential equation (\ref{diffk}) is only sensible if the equation:
\be
h^{M\bar N}\partial_MF\partial_{\bar N}\bar{F} = L(F,\bar{F}) \label{consistency}
\ee  
can be satisfied for some function $L$ of $F,\bar{F}$.  

\section{Calabi-Yau 3-folds from 2-folds}
In this section we lift the D6-brane metric to 11-dimensions and relate this to Calabi-Yau 3-folds.  It is an intricate story.  

The lift of our ten dimensional metric (\ref{g10new}) to 11-dimensions is given by the familiar circle bundle over the 10-d type IIA string-frame metric:
\bea
ds_{11}^2 &=& e^{-{2\phi\over 3}}ds_{10}^2 +e^{{4\phi\over
3}}(d\psi+A_I dx^{I})^2 \\ \nonumber
&=& \eta_{\mu\nu}dx^\mu dx^\nu + H^{-4}dy^2 +H^4(d\psi+A_M dz^{M} + A_{\bar M} dz^{\bar M})^2 \\
&+& 2g_{M\bar N}dz^Mdz^{\bar N} 
\eea
The space-time is a product R$^{(1,4)}\times \cal N$ where $\cal N$ is a 6-dimensional manifold with metric:
\be
ds^2 = H^{-4}dy^2 +H^4(d\psi+A_M dz^{M} + A_{\bar M} dz^{\bar M})^2 \\
+ 2g_{M\bar N}dz^Mdz^{\bar N}. \label{g6}
\ee
Since it preserves 8 supercharges, $\cal N$ must be a Calabi-Yau 3-fold. 

In the remainder of this section we determine the calibrating forms on $\cal N$, show that $\cal N$ is a complex manifold, that it is K{\" a}hler, that there is a global (3,0) form, and determine the topology of $\cal N$.

\subsection{Calibrating forms on $\cal N$}
We expect to have a 2- and a 3-form which calibrate minimal 2 and 3-submanifolds.  These calibrating forms are the K{\" a}hler and (3,0) forms on the 3-fold as we shall show a little later.  For the moment we simply want to show that all the calibrating forms that we have found before can be collected together in these two forms.  

Consider first a D2-brane wrapped on a energy minimizing holomorphic submanifold of the 2-fold.  This lifts to an M2-brane wrapped on a holomorphic 2-cycle of the 3-fold $\cal N$.  The mass of this configuration is give by (\ref{d2}):
\be
m = T_2\int_{\Sigma_2} J' = T_{M2}\int_{\Sigma_2} J'.
\ee
Our method suggests that $J'$ should be closed with respect to the exterior derivative $d_6$ on $\cal N$.  This is too strong a condition, however.  All we need is that it is closed upto a 2-form which has zero integral over a minimal 2-cycle which is completely contained in $\cal M$.  We can calculate the exterior derivative of $J'$ to find:
\be
d_6J' = \partial_y J'\wedge dy + d_4 J' =  \partial_y J'\wedge dy.
\ee
So this is alright since it satisfies our criterion that $J'$ calibrates holomorphic 2-cycles in $\cal M$.  We are interested in finding a calibrating 2-form on $\cal N$ with $J'$ a term in it - since $J'$ calibrates those cycles inside of the 2-fold.  It is easy to see that (\ref{F}) implies:
\be
d_6 J' = F_2\wedge dy.
\ee
and therefore:
\be
d_6(J' + dy\wedge A) =0.
\ee
It might be tempting to identify $J' + dy\wedge A$ as our calibrating 2-form, but it is easy to see that on its own it is not gauge invariant.  Only the combination $d\psi + A$ is gauge invariant.  If we make this substitution we find a gauge invariant quantity that is manifestly closed and gauge invariant:
\be
\omega = J' + dy\wedge (d\psi + A) \label{omega}.
\ee
While the first of the terms in the above expression has a simple interpretation in terms of calibrations the other two were included to make the expression closed under $d_6$ while maintaining gauge invariance.  As we now show, this new term also calibrates brane configurations.  The term $dy\wedge d\psi$ calibrates M2-branes wrapped on the M-theory circle $\psi$ and stretched along $y$, as can be seen by writing the volume form on this space:
\be
H^2dy \wedge (H^{-2})d\psi = dy\wedge d\psi.
\ee  
These correspond to the minimal energy F1-strings stretched between D6-branes that we studied in section 3\footnote{In comparing the expression one must use the relation $T_{F1} = T_{M2}\int d\psi$.}.  From this point of view, $dy\wedge A$ had to be included to make $dy\wedge d\psi$ gauge invariant.  We will show a little later in this section that $\omega$ is the K{\"a}hler form on $\cal N$.  

We turn now to 3-cycles in $\cal N$.  Consider D4-branes wrapped on special Lagrangian 2-cycles on $\cal M$.  D4-branes are M5-branes wrapped on the M-theory circle $\psi$.  The wrapped M5-branes are 2-branes in the flat part of the space time which preserve supersymmetry.  The tension of the 2-brane is given by (\ref{d4}):
\be
T = T_4\int H^2\Omega' =T_{M5}\int H^2\Omega' \wedge d\psi
\ee  
with the understanding that $T_4 = T_{M5}\int d\psi$.  From our previous results we know that $H^2\Omega' \wedge d\psi$ is closed under $d_4$.  We invoke a further fact about Calabi-Yau 3-folds which provides an argument for why $\partial_y (H^2\Omega ')=0$.  The 3-form $H^2\Omega '\wedge d\psi$ should be part of a {\it holomorphic} (3,0) form which calibrates 3-cycles.  If it depended on $y$ it would not be possible to make it part of a holomorphic 3-form without including a term that depended explicitly on $\psi$.  We have already assumed that $\partial_\psi$ is a Killing vector and therefore all quantities are forbidden from having any dependence on $\psi$.  This proves that $\partial_y(H^2\Omega') =0$.  It follows then:
\be
d_6(H^2\Omega'\wedge d\psi) =0.
\ee    
We have thus constructed a closed 3-form.  Again we encounter the problem of non-gauge invariance.  If we substitute $d\psi \rightarrow d\psi + A$ to make the expression gauge invariant, the new expression $H^2\Omega'\wedge (d\psi + A)$ is not closed, in fact:
\be
d_6(H^2\Omega'\wedge (d\psi + A)) = H^2\Omega'\wedge F_2 =i H^2\Omega'\wedge(\partial_{\bar M}H^{-4}dz^{\bar M}\wedge dy)
\ee
In the last equality we used that $J'\wedge \Omega' =0$.  It is possible to construct a new closed gauge invariant 3-form $\zeta$:
\be
\zeta = H^2\Omega'\wedge(d\psi + A -iH^{-4} dy).
\ee
The last term is added to make the 3-form closed.  This new term has a simple interpretation as it is precisely the calibration needed for NS5-branes wrapped on special Lagrangian 2-cycles in $\cal M$ and stretched along the $y$-direction.  This is the calibrating form found earlier in section 3 on minimal cycles with a vanishing pullback of the gauge field $A$.  The 3-form $\zeta$ succinctly combines all the calibrating forms we discovered in section 3 for submanifolds that are 3 dimensional when the geometry is lifted to 11-dimensions.  

If we return to the remaining brane intersections of section 3, one can straightforwardly check that all of the tensions calculated there are reproduced by $\omega$ and $\zeta$.  Thus all the calibrations in 10-d are contained in their 11-d counterparts: $\omega$ and $\zeta$.

We will see shortly that $\omega$ is the K{\" a}hler form and that $\zeta$ is the unique (3,0) form on $\cal N$.  But even at this stage it is possible to prove an important relation between $\omega$ and $\zeta$:
\bea
\omega^3 &=& 3J'\wedge J'\wedge dy\wedge (d\psi + A) \nonumber \\
&=& -\frac{3}{2}\Omega'\wedge \Omega'\wedge dy\wedge (d\phi + A) \nonumber \\
&=& -i\frac{3}{4}\zeta \wedge {\bar\zeta}.
\eea 
This is exactly the relationship that the K{\"a}hler form and the (3,0) form must satisfy.  

\subsection{$\cal N$ is a K{\" a}hler manifold}
There are a number of ways to proceed here.  We follow a route that uses the calibrating forms from the previous section.   We define an almost complex structure on $\cal N$.  Let $e^u, e^v$ be (1,0) one forms with respect to the complex structure on $\cal M$ that provide a local frame:
\bea
g_{M\bar N}' &=& \frac{1}{2}(e^u_Me^{\bar u}_{\bar N} +e^v_Me^{\bar v}_{\bar N}) \nonumber \\
J' &=& \frac{i}{2}(e^u\wedge e^{\bar u} + e^v\wedge e^{\bar v}) \\
\Omega' &=& e^u\wedge e^v.
\eea
On $\cal N$ we define an almost complex structure:
\bea
E^u &=& e^u \\
E^v &=& e^v\\
E^w &=& -iH^{-2}dy + H^2(d\psi + A)
\eea
One can check that with this choice:
\bea
G_{IJ} &=& \frac{1}{2}(E^u_I E^{\bar u}_J + E^{\bar u}_I E^{u}_J + E^v_I E^{\bar v}_J + E^{\bar v}_I E^{v}_J + E^w_I E^{\bar w}_J + E^{\bar w}_I E^{w}_J)\\
\omega &=& \frac{i}{2}(E^u\wedge E^{\bar u} + E^v\wedge E^{\bar v} + E^w\wedge E^{\bar w}) \\
\zeta &=&  E^u\wedge E^v\wedge E^w.
\eea
$G_{IJ}$ in the above is the metric on $\cal N$ given in (\ref{g6}).
Thus, $\omega$ and $\zeta$ are (1,1) and (3,0) forms with respect to this almost complex structure.  We now show that this almost complex structure is integrable.  Recall \cite{chern} that in general a (1,0) form $E^m$ satisfies:
\be
dE^m = a^m_{np} E^n\wedge E^p +  b^m_{n\bar p} E^n\wedge E^{\bar p} + c^m_{\bar n\bar p} E^{\bar n}\wedge E^{\bar p}
\ee
That is, the exterior derivative of a (1,0) form can have terms that are classified as (2,0), (1,1) and (0,2) forms.  The almost complex structure is integrable if and only if there is no (0,2) piece \cite{chern} (i.e. $c^m_{\bar n\bar p}=0$).  In the case of $E^u$ and $E^v$, we have already assumed that there there are complex coordinates in terms of which
\be
E^m = e^m_Mdz^M,
\ee
thus for these two cases the coefficients $c^m_{\bar n\bar p}$ vanish.  In the case of $E^w$ we find:
\bea
dE^w &=& -2E^N_m\partial_N\ln H E^m\wedge E^w + 4E_m^N\partial_N\ln H E^m\wedge E^{\bar w} \nonumber \\
&+& 2 E_{\bar m}^{\bar N}\partial_{\bar N}\ln H E^{\bar m}\wedge E^{w}
 - \frac{i}{2}\partial_y H^2 E^{w}\wedge E^{\bar w} \\&+&i\partial_y g_{M\bar N}'E^M_mE^{\bar N}_{\bar n}E^m\wedge E^{\bar n}. \nonumber
\eea
Since there are no (0,2) terms in the above expression the almost complex structure on $\cal N$ is integrable.  This completes our proof that $\cal N$ is a complex manifold.   

When the almost complex structure is integrable it is possible to define a complex structure or, equivalently, there are local holomorphic coordinates.  We already have $z^M$ defined on $\cal M$ that we inherit as holomorphic coordinates on $\cal N$.  We now construct a third holomorphic differential in terms of which one can define a new complex coordinate.  Let:
\bea
d\eta^1 &=& dz^1, \nonumber \\
d\eta^2 &=& dz^2, \\
d\eta^3 &=& d\psi -iH^{-4}dy  +({\bar a} - a). 
\eea
We have introduced the following notation here
\bea
a &=& -\frac{i}{2}\partial_y \partial_M K dz^M \nonumber \\
A &=& a +{\bar a}
\eea
That is, $a$ is the (1,0) part of $A$.  
It is straightforward to check that $d(d\eta^3)=0$ by using (\ref{neweom}).  In addition:
\be
E^w = 2H^2a_M d\eta^M + H^2d\eta^3
\ee
where $M=1,2$ in the above expression.

In this subsection we showed that $\cal N$ is a complex manifold and that the would-be K{\"a}hler form $\omega$ is closed.  These prove that $\cal N$ is K{\" a}hler.  In addition we showed that $\zeta$ is a (3,0) form.  It is globally well defined since it measures the volume of minimal 3-submanifolds.  

\subsection{The topology of ${\cal N}$}
In addition to the non-trivial cycles inherited by $\cal N$ from the underlying Calabi-Yau 2-fold, there are non-trivial 2-cycles due to the presence of the D6-branes.  To see this, we note that the harmonic function (cf  equation (\ref{harmonic}))$H^{-4}$ blows up at the location of the D6-branes.  One can see this by noting that the source equation $dF_2 =0$ is satisfied everywhere except at the location of the D6-branes where there is a "delta-function" singularity.  This indicates that $H^{-4}$ is singular at the locations of D6-branes.  At the locations of the D6-brane the circle fiber with coordinate $\psi$ collapses since the radius of the fiber is given by $H^4$.  Now consider two separated D6-branes.  Let their locations be labeled by $p_1$ and $p_2$, respectively.  Then, just as in \cite{sen}, we can consider a line joining $p_1$ and $p_2$.  As long as we don't encounter any other D6-branes on this line the circle fiber is non-singular except at the end points.  Thus we have a circle fibered over a line segment with the fiber collapsing only at the ends - it has the topology of a 2-sphere.  The minimal volume of this 2-sphere is measured by $\omega$ as we discussed earlier when the D6-branes were separated along the $y$ direction.  

These 2-cycles can be labeled as $S_{ij}$ where $i,j$ are D6-brane labels.  Each pair of distinct D6-branes is associated with a 2-cycle.  The 2-cycle, $S_{ij}$, say,  collapses when the locations of D6-branes $i$ and $j$ coincide.  Notice that $S_{ij}+S_{jk}= S_{ik}$ in the sense of homology.   We have thus shown that there are an additional $N-1$ 2-cycles relative to the topology of the underlying Calabi-Yau 2-fold.

We now attempt to construct elements of the cohomology group $H^{(1,1)}$ of $\cal N$ corresponding to the proliferation of 2-cycles due to the presence of D6-branes.  To do this, we use our ansatz (\ref{ansatz}) for the geometry which gives us the linear differential equation (\ref{harmonic}).  We write:
\bea
H^{-4} &=& 1+ \sum_{l=1}^{N}h^{(l)} \nonumber \\
K_1 &=& \sum_{l=1}^{N} k^{(l)}.
\eea
with the label $l$ of $h^{(l)}$ and $k^{(l)}$ running over the $N$ D6-branes.
With this ansatz, equation (\ref{neweom}) reduces to $N$ independent equations:
\be
2h^{(l)} + \partial_y^2k^{(l)} =0.
\ee
The gauge field $A$ also splits up into a sum:
\bea
A &=& \sum_l A^{(l)} \\
A^{(l)} &=& -\frac{i}{2}\partial_y\partial_M k^{(l)}dz^M + \frac{i}{2}\partial_y\partial_{\bar M}k^{(l)} dz^{\bar M}
\eea
This split is a consequence of our ansatz for $K_1$ which we introduced for the specific purpose of  making our problem linear.  

Consider the set of differential 2-forms:
\bea
\alpha^{(l)} &=& d(H^4h^{(l)})\wedge (d\psi+A) + H^4h^{(l)}F_2 - dA^{(l)} \nonumber \\
&=&d(H^4h^{(l)}(d\psi+A)  - A^{(l)})
\eea
with $l=1,..., N$.  These 2-forms are manifestly real.  In general, 2-forms are a sum of (2,0), (0,2) and (1,1) forms.  The $\alpha^{(l)}$ are (1,1) forms as can be checked by noting that $\alpha^{(l)}\wedge \zeta =0$.  The $\alpha^{(l)}$ are closed, as is clear from the second equality in the above equation.  Therefore we have constructed a set of $N$ (1,1) closed differential forms.  Our construction is based on a similar construction for Taub-NUT spaces presented in \cite{ruback} (see also \cite{sen}).  It is important to note that these $N$ closed forms are not independent.  Note that
\be
\sum_{l=1}^N\alpha^{(l)} = -d(H^4(d\psi +A))
\ee    
is an exact form.  Therefore, there are only a total of $N-1$ independent (1,1)-forms that are closed but not exact.  

We can conclude that $h^{(1,1)}_{CY_3} = h^{(1,1)}_{CY_2} + N-1$ where $N$ is the number of D6-branes, and $\alpha^{(l)}$ provide the remaining $N-1$ closed (1,1)-forms.  

\section{Some explicit examples}
It would be interesting to construct some explicit examples based on the formalism presented above.  In this section we apply the general formalism to a few cases.  The first case is a reproduction of known results but derived independently, while the second is, as far as I am aware, a new result.

\subsection{Taub-NUT spaces}
In this example we consider the case where the Calabi-Yau 2-fold is simply C$^2$.  If we denote by $z^1$ and $z^2$ the global coordinates on C$^2$, we will "wrap" our D6-branes on the $z^2$-plane.  In other words our D6-branes are wrapped on holomorphic submanifolds given by $z^1 = c_i$.  Ordinarily, we would say that the D6-branes are localized at $z^1 = c_i$.  We have to solve (\ref{neweom}).  Our linearization ansatz tells us that $K_1$ depends only on a single holomorphic function of $z^1, z^2$.  It is clear that the holomorphic function that we want $K_1$ to depend on is $F(z^1, z^2) =z^1$.  We solve the harmonic equation (\ref{harmonic}) that replaces (\ref{neweom}) once we make the linearization ansatz.  Thus (using the metric $h_{M\bar N} = \eta_{M\bar N}$ on C$^2$):
\bea
0&=&\partial_y^2H^{-4} + 2h^{1\bar 1}\partial_1\partial_{\bar 1}H^{-4} \nonumber \\
&=& \partial_y^2H^{-4} + 4\partial_1\partial_{\bar 1}H^{-4}
\eea
This equation can be solved:
\be
H^{-4} = 1 +\sum_{i=1}^N \frac{q}{(|z^1 -c_i|^2 + (y-a_i)^2)^{1/2}}
\ee
where we have allowed the D6-branes to be at independent positions $y=a_i$ in the $y$ direction.  We can also find $K_1$ explicitly through the relation (\ref{neweom}):
\be
K_1 = 2q\sum_{i=1}^N[(y-a_i)\ln\{(y-a_i)+((y-a_i)^2+|z^1-c_i|^2)^{(1/2))}\}-((y-a_i)^2+|z^1-c_i|^2)^{(1/2)}].
\ee
The metric $g'$ is:
\bea
g_{1\bar 1} &=& h_{1\bar 1} + \partial_1\partial_{\bar 1}K_1 = \frac{1}{2}H^{-4} \nonumber \\
g_{2\bar 2} &=&  h_{2\bar 2} + \partial_2\partial_{\bar 2}K_1 = \frac{1}{2}
\eea
The gauge field $A$ can be calculated explicitly using (\ref{A}).  Here we simply write down the expression for $F_2$ in this case using (\ref{F}):
\be
F = idy\wedge (\partial_1 H^{-4} dz^1 - \partial_{\bar 1} H^{-4} dz^{\bar 1}) + \frac{i}{2}\partial_y H^{-4} dz^1\wedge dz^{\bar 1}
\ee
The complete metric on $\cal N$ is given by:
\be
ds^2 = |dz^2|^2 +H^{-4}(dy^2 + |dz^1|^2) + H^{4}(d\psi + A_1 dz^1 + A_{\bar 1}dz^{\bar 1})^2
\ee
The reader will recognize this geometry as that of C$\times TN_N$ where $TN_N$ denotes $N$-centered Taub-NUT space.

\subsection{D6-branes in $A_1$ spaces}
Next we consider wrapping D6-branes on the collapsed cycle of an $A_1$ ALE space.  We begin our analysis by writing the metric on a resolved $A_1$ space.  We choose to write the metric in a form that is inspired by \cite{candelas} because the physical interpretation is clear.

An $A_1$ singularity can be expressed as a submanifold of C$^3$: 
\be
x_1^2 +x_2^2 + x_3^2 =0
\ee
which, after a change of coordinates, can be re-written as:
\be
\alpha\beta - \gamma^2 =0. \label{a1}
\ee
Following \cite{candelas} we replace this equation with:
\be
\matrix{\left(\begin{array}{cc} \alpha & \gamma \\ \gamma& \beta \end{array}\right)}\matrix{\left(\begin{array}{cc} \lambda_1 \\ \lambda_2\end{array}\right)} =0
\ee
In this equation \cite{candelas} $\lambda_1, \lambda_2$ are only defined up to a multiplicative factor.  The pair of equations amount to the previous equation (\ref{a1}) whenever $A,B$ and $C$ are not all zero.  For such points the pair $(\lambda_1, \lambda_2)$ determine a point on $P_1$.  When the entries of the matrix all vanish, $\lambda_1, \lambda_2$ are unrestricted and are coordinates on a whole $P_1\sim S^2$.  The non-trivial 2-cycle has two patches $H_+$ where $\lambda_1\neq 0$ and $H_-$ where $\lambda_2\neq 0$.  On the patch $H_+$, $\lambda =\lambda_2/\lambda_1$ is a good coordinate while on $H_-$, $\mu=\lambda_1/\lambda_2$ is a good coordinate.  On the overlap the two coordinates are related through the holomorphic equation $\lambda = 1/\mu$.   

Consider the metric:
\be
ds_2^2 = 2h_{M\bar N}dz^M dz^{\bar N}
\ee
where $h$ is derived from a K{\"a}hler potential for which we take the ansatz (again inspired by the conifold \cite{candelas}):
\be
K_0 = G(r^2) + a^2\ln (1+|\mu|^2)
\ee
where $G$ is a function of 
\bea
r^2&=& |x_1|^2  + |x_2|^2 + |x_3|^2 \nonumber \\
&=& \frac{1}{2}(|\alpha|^2 + |\beta|^2 + 2|\gamma|^2).
\eea

On the patch $H_-$, $\gamma=-\mu \alpha$, $\beta=\mu^2 \alpha$ and 
\be
r^2 = \frac{1}{2}|\alpha|^2(1+|\mu|^2)^2.
\ee
Defining $G' = \partial_{r^2} F$, the determinant of the metric $h$ can be written as
\be
\sqrt{\det h} = \frac{1}{2}\partial_{r^2}\{(r^2G')^2 + a^2(r^2G')\}.
\ee 
Requiring that the metric is Ricci flat amounts to demanding $\sqrt h = c$ where $c$ is a constant.  The most general solution is:
\be
r^2G' = \frac{1}{2}(-a^2 \pm \sqrt{a^4 + 4(2cr^2+c_0)}) 
\ee
Standard normalization in the limit $a\rightarrow 0$ sets $c=1/2$ and a choice of the plus sign in the square root.  Requiring also that we get the metric on $P_1$ at $r=0$ implies that $c_0=0$.  We can integrate one more time to obtain:
\be
G = \frac{1}{2}[2 \sqrt{a^4+4r^2} -a^2\ln r^2 + a^2\ln\{\frac{\sqrt{a^4+4r^2}-a^2}{\sqrt{a^4+4r^2}+a^2}\}]
\ee
The metric $h$ derived from the above K{\"a}hler potential is that of a resolved $A_1$ singularity, where the non-trivial 2-sphere is located at $r=0$ and has radius $a/2$.  The coordinate on the 2-sphere is $\mu$ on this patch.  The same analysis can be carried out on the patch $H_+$ with $\alpha = \lambda^2\beta, \gamma = -\lambda\beta$.    

We wish to solve for the metric of a D6-brane wrapped on the blown-up two cycle of an $A_1$ ALE space.  The coordinate on the sphere on the patch $H_-$ is $\mu$.  This would be a good coordinate on the wrapped D6-brane.   We wish to solve (\ref{harmonic}) with the metric $h$ as above.  To do so we must find a holomorphic function $F$ on the resolved ALE space that satisfies (\ref{consistency}) .  The holomorphic function $F$ serves as a holomorphic coordinate transverse to the brane.  Unfortunately we can only find such a function when $a=0$, the case when the two-cycle is collapsed.  This may simply be a failure of imagination on our part or it could suggest that the linearization we performed in section 5 fails in this case.  In any case we show how to find the metric when $a=0$.  We work in the patch $H_-$ where $\mu$ is a good coordinate.  We pick $F = \alpha$.  The 2-cycle is located at $\alpha=0$.  In this ($a=0$) case the metric takes the simple form:
\bea
h_{\alpha\bar \alpha}&=& \frac{1}{2\sqrt 2}\frac{1+|\mu|^2}{|\alpha|}  \nonumber \\
h_{\alpha\bar \mu}&=&\frac{1}{\sqrt 2}\mu (\frac{\bar \alpha}{\alpha})^{1/2} \\
h_{\mu\bar \mu}&=&{\sqrt 2}|\alpha | \nonumber
\eea
The harmonic equation (\ref{harmonic}) takes the form
\be
0 = (\partial_y^2 + 4\sqrt{2}|\alpha|\partial_\alpha\partial_{\bar \alpha})H^{-4}(y,\alpha,{\bar \alpha})
\ee
This equation can be solved for a stack of branes at $y=y_i$ and $\alpha=0$:
\be
H^{-4} = 1 + q\sum_{i} \frac{1}{((y-y_i)^2 + 2^{3/2}|\alpha|)^{1/2}}
\ee
The K{\"a}hler potential $K_1$ can also be found using equation (\ref{neweom}):
\be
K_1 = q\sum_{i=1}^N(2\sqrt{(y-y_i)^2+ 2^{3/2}|\alpha|} - (y-y_i)\ln \frac{(y-y_i)+\sqrt{(y-y_i)^2+ 2^{3/2}|\alpha|}}{(y-y_i)-\sqrt{(y-y_i)^2+ 2^{3/2}|\alpha|}})
\ee
The Calabi-Yau 3-fold metric (\ref{g6}) is given in terms of $H^{-4}$, the metric $g'$ and the 1-form $A$.  The metric $g_{M\bar N}'= \partial_M\partial_{\bar N}(K_0+K_1)$ is given by:
\bea
g_{\alpha\bar\alpha}' &=& h_{\alpha\bar\alpha} +\frac{q}{2^{3/2}|\alpha|}\sum_{i} \frac{1}{((y-y_i)^2 + 2^{3/2}|\alpha|)^{1/2}}\nonumber \\
&=&\frac{1}{2\sqrt 2}\frac{1+|\mu|^2}{|\alpha|} +\frac{q}{2^{3/2}|\alpha|}\sum_{i} \frac{1}{((y-y_i)^2 + 2^{3/2}|\alpha|)^{1/2}}  \nonumber\\
g_{\alpha\bar \mu}' &=& h_{\alpha\bar \mu} = \frac{1}{\sqrt 2}\mu (\frac{\bar \alpha}{\alpha})^{1/2} \\
g_{\mu\bar \mu}'&=&h_{\mu\bar \mu}={\sqrt 2}|\alpha | \nonumber
\eea
while the 1-form $A$ is:
\be
A = -\frac{iq}{2}\sum_{i=1}^N\frac{(y-y_i)}{\sqrt{ (y-y_i)^2+ 2^{3/2}|\alpha |}}(\frac{1}{\alpha}d\alpha - \frac{1}{\bar\alpha}d{\bar \alpha})
\ee
The metric for the Calabi-Yau 3-fold (\ref{g6}) is completely specified in terms of the above quantities.  

We can also construct the (3,0) form and K{\"a}hler form for the 3-fold.  The (3,0) form is expressed in terms of the (2,0) form $\Omega'$.  In the coordinates $\alpha, \mu$ the (2,0) form takes a very simple form:
\be
\Omega '= {\sqrt 2}H^{-2}d\alpha\wedge d\mu
\ee
the (3,0) form is then:
\be
\zeta = {\sqrt 2} d\alpha\wedge d\mu\wedge (d\psi + A -iH^{-4}dy) 
\ee
and the K{\" a}hler form $\omega$ is given by the expression:
\bea
\omega &=& J' + dy\wedge (d\psi + A) \nonumber \\
&=& i(\frac{1}{2\sqrt 2}\frac{1+|\mu|^2}{|\alpha|} +\frac{q}{2^{3/2}|\alpha|}\sum_{i} \frac{1}{((y-y_i)^2 + 2^{3/2}|\alpha|)^{1/2}})d\alpha\wedge d{\bar\alpha} \nonumber \\
 &+&  \frac{i}{\sqrt 2}\mu (\frac{\bar \alpha}{\alpha})^{1/2}d\alpha\wedge d{\bar\mu} 
+  \frac{i}{\sqrt 2}{\bar\mu} (\frac{\alpha}{\bar\alpha})^{1/2}d{\mu}\wedge d{\bar\alpha} + 
i{\sqrt 2}|\alpha| d\mu\wedge d{\bar\mu} \\
&+& dy\wedge (d\psi  -\frac{iq}{2}\sum_{i=1}^N\frac{(y-y_i)}{\sqrt{ (y-y_i)^2+ 2^{3/2}|\alpha |}}(\frac{1}{\alpha}d\alpha - \frac{1}{\bar\alpha}d{\bar \alpha})). \nonumber
\eea
 
In addition to the quantities computed above, we can also write down explicit expressions for the additional elements of H$^{(1,1)}$ as explained in section 6.  This completes our discussion of wrapped D6-branes on an $A_1$ singularity.  This wrapped D6-brane geometry is studied from a different point of view in \cite{enunez}, it would be interesting to understand how these two points of view are related to each other more explicitly.  

\section{Conclusions}
This paper provides a supergravity description of D6-branes wrapped on 2-cycles in Calabi-Yau 2-folds.  The supergravity description of the wrapped branes is then related to the geometry of non-compact Calabi-Yau 3-folds through an 11-d lift.  In addition to the metric, the K{\" a}hler form and (3,0) form are explicitly constructed.   

In passing from the general description to explicit solutions one must solve the non-linear differential equation (\ref{neweom}).  We argue that one should be able to write an ansatz that makes this equation linear because in principle one can stack an arbitrary number of D6-branes.  One way to achieve this linearization is explained.  This yields the simpler equation (\ref{harmonic}).  The linearization ansatz is used to construct some explicit examples in section 7.  

In general the 3-folds studied here are argued to have an additional set of $(N-1)$ 2-cycles relative to the ones inherited from the underlying 2-fold.  The additional elements of the cohomology group $H^{(1,1)}$ corresponding to these new cycles are constructed in section 6 using the linearization ansatz.  

There are many directions opened up by our analysis.  The most obvious is to find new solutions to the equations presented here.  These include generalizing the wrapped brane configuration on the singular $A_1$ space presented in section 7 to the resolved case.  

The D6-brane geometries studied here are dual to 5-d gauge theories that preserve 8 supercharges.  These gauge theories are closely related to Seiberg-Witten theory in 4 dimensions.  The study of the 5-d theory was initiated by Nekrasov in \cite{nekrasov}.  It would be interesting to apply the ideas of AdS/CFT    to 5d gauge theories with the help of these geometries.
\newpage
\no
{\Large \bf Acknowledgements}\\

\no
I would like to thank the physics departments at Harvard University and Stockholm University for hospitality during the course of this work and the Swedish Vetenskapsr{\aa}det (VR) for travel funds.  The work presented here grew out of ideas presented in \cite{fh, fh2}.  I am grateful to Tasneem Zehra Husain for an enjoyable collaboration on those papers.  I would like to thank Douglas Smith for his thoughtful comments on an earlier draft of this paper.  I am also grateful to Jos{\' e} Edelstein and Carlos N{\' u}{\~n}ez for sharing their insights on this problem.

}

\begin{thebibliography}{77}
%
\bibitem{fh}
A.~Fayyazuddin and T.~Z.Husain {\it Calibrations, Torsion Classes, and Wrapped M-branes } hep-th/0512030.
%
\bibitem{fh2}A.~Fayyazuddin and T.~Z. Husain {\it $G_2$ holonomy metrics and wrapped D6-branes} hep-th/0608163.

%
\bibitem{gm}
J.~Gomis, {\it D-Branes, Holonomy and M-Theory }, hep-th/0103115

%
\bibitem{enunez}
J.~ D.~Edelstein and C.~N{\' u}{\~n}ez, {\it D6 branes and M-Theory geometrical transitions from gauged supergravity} hep-th/0103167

%
\bibitem{martucci}
L.~Martucci and P.~Smyth, {\it Supersymmetric D-branes and calibrations on general N = 1
backgrounds}, hep-th/0507099.\\
L.~Martucci,
{\it D-branes on general N = 1 backgrounds: Superpotentials and D-terms},
hep-th/0602129.  

%
\bibitem{gencal}
J.~Gutowski and G.~Papadopolous {\it AdS Calibrations}, hep-th/9902034 \\
J.~Gutowski, G.~Papadopolous and P. ~K.~Townsend {\it Supersymmetry and Generalised Calibrations}, hep-th/9905156\\
H.~Cho, M.~Emam, D.~Kastor and J.~Traschen
{\it Calibrations and Fayyazuddin-Smith Spacetimes}
hep-th/0009062

%
\bibitem{chern}
S. S. Chern {\underline Complex manifolds without potential theory}, 2nd Edition, Springer-Verlag, New York, 1979.

%
\bibitem{sen}
A. Sen {\it A Note on Enhanced Gauge Symmetries in M- and String Theory}, hep-th/9707123.

%
\bibitem{ruback} 
P. Ruback, {\it The Motion of Kaluza-Klein Monopoles}, Comm. Math. Phys. {\bf 107} (1986) 93.

%
\bibitem{candelas}
P. Candelas and X. C. de la Ossa {\it Comments on Conifolds}, Nuc. Phys. {\bf B342} (1990) 246.

%
\bibitem{nekrasov}
N. Nekrasov {\it Five Dimensional Gauge Theories and Relativistic Integrable Systems}, hep-th/9609219.
\end{thebibliography}
\end{document}